# Synthesis, properties and thermal decomposition of the Ta$_4$AlC$_3$ MAX phase

Matteo Griseri[a,b] *, Bensu Tunca [a,b], Thomas Lapauw[a], Shuigen Huang[a], Lucia Popescu[b], Michel W. Barsoum[c], Konstantina Lambrinou[b], Jozef Vleugels[a]

[a] KU Leuven, Department of Materials Engineering, Kasteelpark Arenberg 44, 3001 Leuven, Belgium
[b] SCK•CEN, Boeretang 200, 2400 Mol, Belgium
[c] Department of Materials Science and Engineering, Drexel University, Philadelphia, PA 19104, USA
* Corresponding author: KU Leuven, Department of Materials Engineering, Kasteelpark Arenberg 44, 3001 Leuven, Belgium.
E-mail address: matteo.griseri@kuleuven.be (M. Griseri).

**Abstract**
The present work describes a synthesis route for bulk Ta$_4$AlC$_3$ MAX phase ceramics with high phase purity. Pressure-assisted densification was achieved by both hot pressing and spark plasma sintering of Ta$_2$H, Al and C powder mixtures in the 1200-1650°C range. The phases present and microstructures were characterized as a function of the sintering temperature by X-ray diffraction and scanning electron microscopy. High-purity α-Ta$_4$AlC$_3$ was obtained by hot pressing at 1500°C for 30 min at 30 MPa. The β-Ta$_4$AlC$_3$ allotrope was observed in the samples produced by SPS. The Young's modulus, Vickers hardness, flexural strength and single-edge V-notch beam fracture toughness of the high-purity bulk sample were determined. The thermal decomposition of Ta$_4$AlC$_3$ into TaC$_x$ and Al vapour in high (~10$^{-5}$ mbar) vacuum at 1200°C and 1250°C was also investigated, as a possible processing route to produce porous TaC$_x$ components.

**Keywords:** MAX Phases; Ta$_4$AlC$_3$; Hot Pressing; Spark Plasma Sintering; Thermal Decomposition

# 1. Introduction

The M$_{n+1}$AX$_n$ (MAX) phases are layered ternary ceramic compounds, where M is an early transition metal, A is a metal from groups 13-15 in the periodic table of elements, X is carbon or nitrogen, and n = 1, 2 or 3 [1]. The MAX phases crystallize in the hexagonal system, space group *P6$_3$/mmc*, and exhibit compound-specific hybrid ceramic/metallic properties, such as refractoriness, oxidation/corrosion resistance, damage tolerance, machinability, electrical/thermal conductivity, etc. [1]. The synthesis of MAX phase-based ceramics can be realized by reactive sintering of the constituent elemental powders [2,3]; however, the achievement of high phase purity is often challenging due to the concurrent formation of intermetallic and binary carbide phases. The present work studies the production and decomposition in vacuum of the not so extensively studied MAX phase Ta$_4$AlC$_3$. In previous works, small Ta$_3$AlC$_2$ and Ta$_4$AlC$_3$ single crystals were grown via a molten metal technique [4]. Producing bulk, polycrystalline MAX phase ceramics, however, typically relies on powder metallurgical routes that involve pressure-assisted densification methods, such as hot pressing (HP) or spark plasma sintering (SPS). SPS relies on local direct volumetric Joule heating by driving an electrical current through the powder compact, as opposed to HP that uses an external heat source [3]. SPS is preferred over HP due to its direct volumetric heating, high heating and cooling rates and lower energy consumption. Moreover, SPS allows the production of fine-grained ceramics, due to the short processing times. Elemental Ta-Al-C powder mixtures have been previously spark plasma sintered or hot pressed into dense polycrystalline samples of $Ta_2AlC$ [5] and $Ta_4AlC_3$ [6].

The MAX phases are known to undergo decomposition into their respective binary MX phases by dissociating the A element at compound-specific temperatures and in various atmospheres [7–9]. To date, the thermal decomposition of MAX phases in diverse atmospheres has been investigated for a limited set of MAX phases, i.e., $Ti_4AlN_3$ [7,10], $Ti_3SiC_2$ [8,9,11–16], $Ti_3AlC_2$ [8,13,15], $Ti_2AlC$ [8,13,15], $Ti_2AlN$ [7,15] and $Cr_2AlC$ [8]. In most cases, the decomposition resulted in the formation of the respective binary phases, e.g., $TiC_x$, $TiN_x$ and $CrC_x$. Of interest to this work is the in-situ thermal decomposition of $Ta_2AlC$, which was studied in argon atmosphere [17]. In the absence of reactive gaseous species, such as $O_2$ or $N_2$, these studies suggest a simple $M_{n+1}AX_n \rightarrow M_{n+1}X + A_n$ decomposition scheme. The major driving force leading to such thermal decomposition is the high vapour pressure of the A element, which diffuses out of the MAX phase grains, evaporating above a certain threshold temperature. Other factors, such as material geometry, phase purity and type of heating used, influence the decomposition kinetics. In bulk $Ti_3SiC_2$, the kinetics of A-element diffusion and evaporation was studied by measuring the depth of the degraded layer as function of temperature and time, while the produced $TiC_x$ was reported to be porous [14].

The interest in $Ta_4AlC_3$ and its thermal decomposition into $TaC_x$ and Al in vacuum lies in the fact that such route can be exploited to produce porous $TaC_x$ components with complex geometries for specific high-temperature applications. One may start by shaping the machineable MAX phase precursor followed by its thermal decomposition. A possible application of this approach is found in the field of radioactive ion beam production, and more specifically in the production of target materials for Isotope Separators On-Line (ISOL). The ISOL technique allows the on-line evaporation, selective ionization, acceleration and mass-purification of radioisotopes that are produced in a target material irradiated by an accelerated particle beam [18,19]. Materials for ISOL targetry should combine the necessary refractoriness to withstand the high-power beam deposition with an optimised microporosity that enables a fast evaporation of the requisite reaction products. Therefore, a microporous binary carbide ceramic produced by the thermal decomposition of a MAX phase ternary carbide, such as $Ta_4AlC_3$, may provide an innovative and promising target material for ISOL facilities.

In summary, present work describes the synthesis of high-purity $Ta_4AlC_3$ by reactive HP and SPS from $Ta_2H$-Al-C powder mixtures. The two pressure-assisted densification methods were compared in terms of the phase assembly, microstructure, density and mechanical properties of the produced bulk $Ta_4AlC_3$ MAX phase ceramics. The phase purity and microstructure of these ceramics were characterized by means of X-ray diffraction (XRD), scanning electron microscopy (SEM), and energy-dispersive X-ray spectroscopy (EDS), while their Young's modulus, Vickers hardness, 4-point bending flexural strength, and single-edge V-notch beam (SEVNB) fracture toughness were also determined. Moreover, the thermal decomposition of the purest $Ta_4AlC_3$ grade was studied at 1200°C and 1250°C in vacuum, and the resulting products were characterized by SEM, EDS, and transmission electron microscopy (TEM).

## 2. Experimental procedure
### 2.1. Material synthesis

Initial attempts to produce $Ta_4AlC_3$ by reactive HP starting with Ta (particle size <80 μm, purity 99.95%, H.C. Starck) Al (particle size <5 μm, >99% purity, AEE, USA) and C (particle size <5 μm, >99% purity, Asbury Graphite Mills, USA) elemental powder mixtures resulted in low phase purity samples, presumably due to inhomogeneities associated with the relatively coarse Ta powder used. In order to optimize the mixing and the contact between particles in the powder compact, the particle size of the as-received Ta powder was reduced by converting it into to a brittle $Ta_2H$ powder prior to ball milling. Powder embrittlement was achieved by heating the starting Ta powder in pure hydrogen (H2) at 800°C for 2 h. The resulting $Ta_2H$ powder was ball-milled (Retsch PM4-MA, Germany) at 250 rpm for 30 min in isopropanol, using 3Y-TZP zirconia ($ZrO_2$) balls (∅ 5 mm and 10 mm) in a 3Y-TZP $ZrO_2$ container. This milling step reduced the maximum particle size from 80 µm to 10 µm. The milled mixture was dried in a rotating evaporator (Heidolph 4010) and sieved through a 32 µm fabric sieve to remove the larger particles.

Milled $Ta_2H$, Al and graphite were dry-mixed in Ar using a multidirectional mixer (Turbula T2C. WAB, Switzerland) for 24 h. The loss of liquid Al by exudation through the punch/die clearance during pressure-assisted sintering was compensated by a 25% hyper-stoichiometric excess of Al. Moreover, a 13% sub-stoichiometric C content compensated for the carbon uptake from the graphite die/punch set-up and to favour the stability of $Ta_4AlC_3$ formation over binary $TaC_x$. Thus, the Ta:Al:C ratio in the starting powder mixture was 4:1.25:2.6. The corrections for both the Al and C contents in the starting powder mixture aimed at inhibiting the formation of the competing phases (binary carbides, intermetallics) in favour of MAX phase formation, as has been previously reported for other Al-containing MAX phases [2,20].

The powder mixture was cold-pressed at 30 MPa into 2 cm-diameter discs and placed in a graphite die. Graphite foil was used as release agent between powder and die wall. The powder compacts were reaction-sintered by SPS (HP D 25, FCT Systeme, Frankenblick, Germany) or HP (W100/150-2200-50 LAX, FCT Systeme, Frankenblick, Germany) in the 1200-1650°C range, under a pressure of 30 MPa, in 0.4 mbar vacuum. The applied heating rate and dwell time at the sintering temperature were 20°C/min and 30 min for HP, and 100°C/min and 15 min for SPS. In both sintering routes, a minimum load of 15.7 MPa was applied to the powder compact during heating, primarily so as to guarantee electrical contact during SPS. The continuous monitoring of the SPS punch displacement over time allowed us to follow the densification process. The herein reported data were corrected for the thermal expansion of the graphite die/punch setup.

## 2.2. Thermal decomposition in vacuum

The dense, bulk, polycrystalline $Ta_4AlC_3$ samples were sliced by wire electrical discharge machining (EDM) into discs with 0.5 - 1 mm thickness. The specimen surfaces were then polished, and placed in an alumina, $Al_2O_3$, crucible for high-temperature annealing in a resistively heated furnace (Brew) with W heating elements. All treatments were conducted in a vacuum of ~$10^{-5}$ mbar. The annealing temperatures were 1200°C and 1250°C, with dwell times varying between 30 and 1980 min (33 h).

## 2.3. Material characterization

The phase assembly of the produced samples (both in bulk and milled powder form) was characterized by XRD in the 5-75° $2\theta$ range in steps of 0.01° at 0.2 s/step (Cu $K_\alpha$ source operating at 40 kV and 30 mA, Bruker D2 Phaser, Germany). The surface layer of the as-sintered discs was removed by grinding prior to XRD analysis. Phase identification was performed using the X'Pert software. Calculations of the predicted patterns and Rietveld refinement were performed using the Materials Analysis Using Diffraction (MAUD) software [21]. The microstructure of select samples was examined by means of SEM (XL30-FEG, FEI, Netherlands). Metallographic cross-sections for SEM analysis were polished to a mirror finish, using progressively finer diamond suspensions from 15 µm to 0.1 µm.

The heat-treated $Ta_4AlC_3$ discs were cross-sectioned and polished in the same way to investigate their decomposition. Elemental analysis by EDS and wavelength-dispersive X-ray spectroscopy (WDS) were performed on an electron probe microanalyser (EPMA; JXA-8530F, JEOL, Japan). TEM (ARM200F, JEOL, Japan) was used to examine specific areas of interest in the porous samples by means of bright-field imaging, selected area electron diffraction (SAED), scanning transmission electron microscopy (STEM) and high-angle annular dark-field (HAADF) analysis. TEM thin foils were lifted out from selected material areas in a focused ion beam (FIB; Nova NanoLab 600 DualBeam, FEI, USA) facility equipped with an electron backscatter diffraction (EBSD) detector.

The Vickers hardness ($HV_{30}$) of bulk $Ta_4AlC_3$ was measured using an indentation load of 30 kg (FV-700, Future-Tech Corp., Japan). The room temperature (RT) flexural strength was measured by 4-point bending on 5 rectangular bars ($45\times4\times3$ mm$^3$) with inner and outer span widths of 20 and 40 mm, respectively (Instron Testing Systems 4467). The SEVNB fracture toughness, $K_{Ic}$, was determined, using the maximum load defined by 3-point bending of 5 rectangular bars ($25\times2.5\times2$ mm$^3$) with a span of 20 mm and a 0.8 mm V-notch (Instron Testing Systems 4467) [22,23]. The Young's modulus was measured at RT by means of the impulse excitation technique (IET; IMCE system, Belgium), according to ASTM standard E1876-08 [24].

## 3. Results and discussion
### 3.1. Phase assembly

The XRD patterns in Fig. 1 show that no Ta-Al-C MAX phases formed in the 1200-1300°C range. Instead, $AlTa_x$ (non-stoichiometric) intermetallic compounds, such as $AlTa_3$, were identified. In the same temperature range, $TaC_x$ binary carbides with varying stoichiometries and x ≈ 0.5-1 were observed. The $Ta_2AlC$ MAX phase formed between 1300°C and 1500°C during HP, while the $Ta_4AlC_3$ phase only became predominant above 1500°C.

After SPS, $Ta_2AlC$ was still present at 1500°C, disappearing completely at 1600°C and forming $Ta_4AlC_3$ in both the α- and β-$Ta_4AlC_3$ polymorphs. The highest relative fraction of α-$Ta_4AlC_3$ was produced by HPing above 1500°C. The following reactions summarize the formation and decomposition of $Ta_4AlC_3$ as function of temperature, i.e., the early (1) and high-temperature (6) formation of binary $TaC_x$, the formation of 211 (2) and α-413 (3,4), the β-413 formation (5), and the 413 phase decomposition into sub-stoichiometric $TaC_x$ (7):

$Ta + xC \rightarrow TaC_x$ (up to 1300°C) (1)
$AlTa_{x\leq1} + (2 - x)TaC_{1/(2-x)} \rightarrow Ta_2AlC$ (up to 1300°C) (2)
$AlTa_{x\leq1} + (4 - x)TaC_{3/(4-x)} \rightarrow \alpha\text{-}Ta_4AlC_3$ (up to 1300°C) (3)
$Ta_2AlC + 2TaC \rightarrow \alpha\text{-}Ta_4AlC_3$ (up to 1400°C) (4)
$\alpha\text{-}Ta_4AlC_3 \rightarrow \beta\text{-}Ta_4AlC_3$ (SPS above 1600°C; phase transformation) (5)
$Ta + xC \rightarrow TaC_x$ (up to 1650°C; competing with the formation of α,β-$Ta_4AlC_3$) (6)
$\alpha,\beta\text{-}Ta_4AlC_3 \rightarrow 4TaC_{0.75} + Al$ (SPS at 1650°C; Al dissociation) (7)

Ab initio calculations of the Gibbs free energy of formation of the 413 polymorphs have predicted a spontaneous transition of α-$Ta_4AlC_3$ to β-$Ta_4AlC_3$ around 1600°C [25]. However, this claim was a point of debate after conflicting results were obtained by independent first principle calculations [26]. In the present work, this transformation was indeed observed in the specimens prepared by SPS, with the coexistence of the two phases at 1600°C and the complete conversion of α- to β-$Ta_4AlC_3$ at 1650°C. The MAX phases are destabilised above compound-specific temperatures, as the A-element dissociates from the ternary carbide system and binary MX phases are, in general, thermodynamically more stable [13]. Reaching temperatures above 1600°C fast enough in the powder compact will produce materials where the competing binary MX phases, such as $TaC_x$ in this case, are predominant, as not enough reaction time is provided in the temperature window where MAX phase formation is favoured. Such a phenomenon was observed in the samples produced by SPS at 1650°C, where $TaC_x$ dominated over both α- and β-$Ta_4AlC_3$. Low signals of β-$Ta_4AlC_3$ phase were also detected in the samples produced by HP at 1600°C, but the relative fraction with respect to α-$Ta_4AlC_3$ phase was small.

Previous experimental studies failed to confirm the theoretically predicted α→β allotropic phase transformation at 1600°C, reporting that the MAX phases in the Ta-Al-C system tend to dissociate around this temperature. At the even higher temperature of 1750°C, no MAX phase was observed, but $TaC_x$ formed instead [27]. However, the present study confirms that the α→β transition effectively occurs in a limited temperature window and after short dwell times, indicating a relative narrow range of temperature stability of the β polymorph. The predominance of $TaC_x$ over $Ta_4AlC_3$ occurs at temperatures as low as 1600°C, after a dwell time of 15 min; therefore, the absence of either α-$Ta_4AlC_3$ or β-$Ta_4AlC_3$ after a 4 h-treatment at 1750°C [27] is no surprise. The fact that a substantial amount of the β-$Ta_4AlC_3$ phase was only observed in samples that underwent fast cooling in the SPS suggests that this phase is only stable in this elevated temperature window, and that a transition to α-$Ta_4AlC_3$ will occur at temperatures below 1600°C.

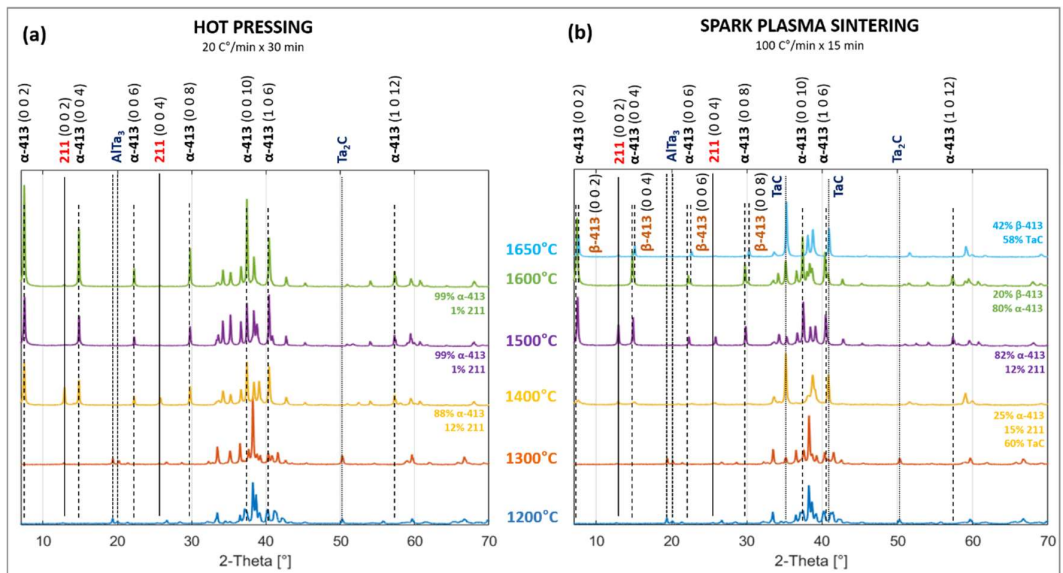

**Fig. 1.** XRD patterns of $Ta_4AlC_3$ samples produced at different temperatures by (a) HP, and (b) SPS. The uncertainty of the phase amounts (in wt%) obtained by Rietveld analysis and shown below each diffractogram is ±2%.

Differences in the sequence of phase formation were observed when comparing the HP and SPS routes. For example, α-$Ta_4AlC_3$ formed from $Ta_2AlC$ at a higher temperature during SPS as compared to HP (Fig. 1), a fact that can be mainly attributed to the faster heating rate and the shorter dwell times during SPS. This suggests that the use of SPS to synthesise bulk $Ta_4AlC_3$ has no distinct advantage in terms of speed over HP, in the case processing aims at the synthesis of high-purity α-$Ta_4AlC_3$. The choice of the current used during SPS could also affect phase transitions, including the formation of β-$Ta_4AlC_3$. Unlike prior work on single crystals [4], this study provided no evidence for the formation of a $Ta_3AlC_2$ MAX phase.

The lattice parameters *a* and *c* of the $Ta_2AlC$, α-$Ta_4AlC_3$ and β-$Ta_4AlC_3$ MAX phases were determined by XRD on powder samples, so as to eliminate micro-strains established during cooling of the bulk samples from their sintering temperatures. Fig. 2 shows the measured and XRD computed patterns of $Ta_2AlC$, α,β-$Ta_4AlC_3$ and $TaC_x$ after Rietveld refinement using the MAUD software [21]. The measured lattice parameters *a* and *c* of the Ta-Al-C MAX phases are compared to data reported for single crystals in Table 1, showing good agreement.

**Table 1**
Lattice parameters of Ta-Al-C MAX phases reported in literature and measured in this work.

| Phase | Single-crystal data [1] | | Measured data | |
| --- | --- | --- | --- | --- |
| | *a* (Å) | *c* (Å) | *a* (Å) | *c* (Å) |
| α-Ta$_4$AlC$_3$ | 3.11 | 24.1 | 3.113(6) | 24.10(9) |
| β-Ta$_4$AlC$_3$ | 3.09 | 23.71 | 3.086(5) | 23.74(1) |
| Ta$_2$AlC | 3.07 | 13.8 | 3.096(8) | 13.90(5) |

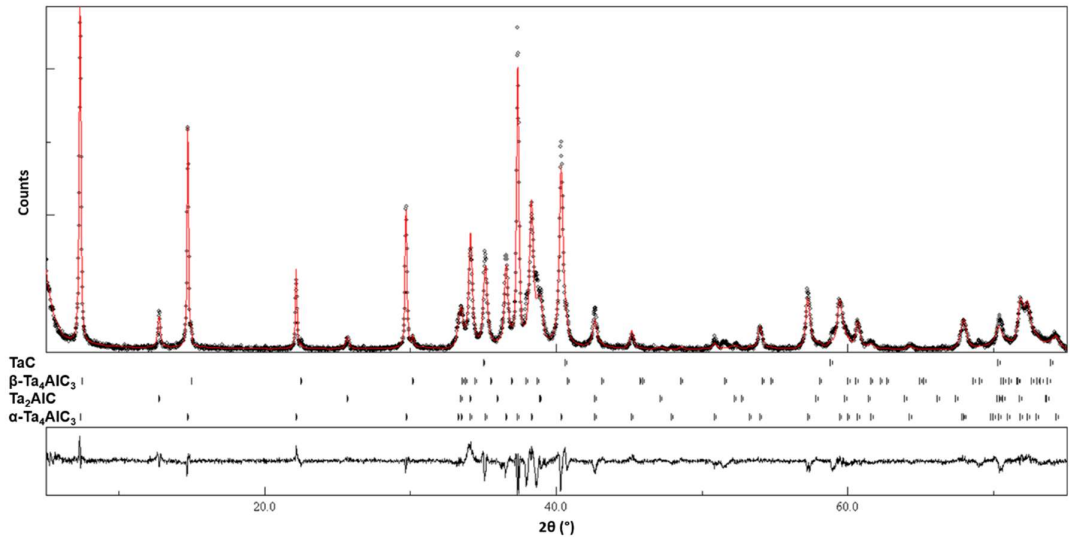

**Fig. 2.** Powder XRD pattern of a sample containing Ta$_2$AlC (9.4wt%), α-Ta$_4$AlC$_3$ (86.0wt%), β-Ta$_4$AlC$_3$ (2.6wt%) and TaC$_x$ (2.0wt%) with identified phase peaks. The computed profile of these four phases was fitted to the experimental pattern by Rietveld refinement using the MAUD software (Rwp = 10%).

### 3.2. Densification

The densities of Ta$_4$AlC$_3$-based samples with high MAX phase content, i.e., those sintered above 1500°C, were measured by the Archimedes' method and the results are listed in Table 2. Since α-Ta$_4$AlC$_3$ was the only phase present in samples made by HP at 1500°C and 1600°C, the measured density can be compared to the theoretical density (TD) of α-Ta$_4$AlC$_3$ (TD = 12.92 g/cm$^3$ [1]) or mixtures of α-Ta$_4$AlC$_3$, β-Ta$_4$AlC$_3$ (TD = 13.36 g/cm$^3$ [1]), Ta$_2$AlC (TD = 11.82 g/cm$^3$ [1]) and TaC (TD = 14.49 g/cm$^3$ [28]). The densities of the samples produced by HP at 1500°C and by SPS at 1600°C were all lower than the theoretical density of α-Ta$_4$AlC$_3$ (Table 2), due to the presence of limited residual porosity and a roughly estimated 1 ± 2 wt% amount of Al$_2$O$_3$ (TD = 3.98 g/cm$^3$ [29]). The weight percent of Al$_2$O$_3$ could not be accurately estimated by Rietveld refinement, because its amount was close to the XRD detection limit and several diffraction peaks overlapped with those of the MAX phases, i.e., planes (012) and (008) in 413 and (004) in 211. Samples produced by SPS at 1600°C and 1650°C had slightly higher densities, as they contained β-Ta$_4$AlC$_3$ and TaC.

**Table 2**
Density of selected grades, as determined by Archimedes' method; TD refers to theoretical density; 211: $Ta_2AlC$; 413: $Ta_4AlC_3$.

| Sintering method | Sintering temp. (°C) | Load (MPa) | Density (g/cm³) | Constituent phases (in wt%) |
|---|---|---|---|---|
| HP | 1500 | 30 | 12.09 (93.6% TD) | α-413 (99), 211 (1), $Al_2O_3$ |
| HP | 1600 | 30 | 12.27 (95% TD) | α-413 (99), 211 (1), $Al_2O_3$ |
| SPS | 1500 | 30 | 11.93 (93.9% TD) | α-413 (80), 211 (20), $Al_2O_3$ |
| SPS | 1600 | 30 | 12.56 (96.5% TD) | α-413 (80), β-413 (20), $Al_2O_3$ |
| SPS | 1650 | 30 | 12.97 (92.5% TD) | TaC (58), β-413 (42), $Al_2O_3$ |

The densification process was monitored in the SPS facility by measuring the relative axial displacement of the vertical graphite punches. Fig. 3 shows the time-displacement curve (indicative of the powder compact shrinkage) of the upper ram (the bottom ram was fixed), along with the temperature cycle and the vacuum level for a sample sintered at 1600°C under an applied load of 30 MPa. As may be seen, the initially applied load corresponds to a pressure of 15.7 MPa. After 120 s, the temperature reached 400°C and the pyrometer started measuring the temperature of the powder compact. From that point onward, the heating rate was temperature-controlled and fixed to 100°C/min. The powder compact underwent a slight thermal expansion during heating up to about 550°C. Above the melting point of Al (660°C), densification was clearly initiated. The sudden drop in the vacuum level around 700°C could be associated with the dehydrogenation of $Ta_2H$. Densification was accelerated around 1300°C, at which point the pressure increase could probably be attributed to the late degassing promoted by densification. The produced gases could be Al vapour and $CO_2$ resulting from the carbothermal reduction of metallic oxides present in the starting powder mixture. The abrupt drop in punch displacement after 850 s corresponds to the application of an increased load up to 30 MPa. Under this load, further densification of the powder compact was achieved. After 15 min of dwell, the load was removed and the temperature decreased quickly due to the interrupted electrical current.

### 3.3. Microstructure

Figs. 4a and b show typical SEM micrographs of polished samples produced by HP and SPS at 1500°C, respectively, revealing a typical MAX phase microstructure with randomly oriented lamellar grains. The average length and width of the grains of a sample produced at 1500 °C by HP (Fig. 4a) were measured to be 6 ± 1 μm and 1.7 ± 1 μm, respectively. In the SPS sample (Fig. 4b), the average length and width were 4.8 ± 1 μm and 1.2 ± 1 μm, respectively, most likely due to the shorter sintering time. An earlier study on $Ta_4AlC_3$ reported larger grains with average length and width of 10 μm and 3 μm, respectively, after 1 h of hot pressing [6]. Differences in grain size are attributed to differences in the sintering cycle duration: longer cycles allow for grain growth, shorter cycles produce fine-grained samples.

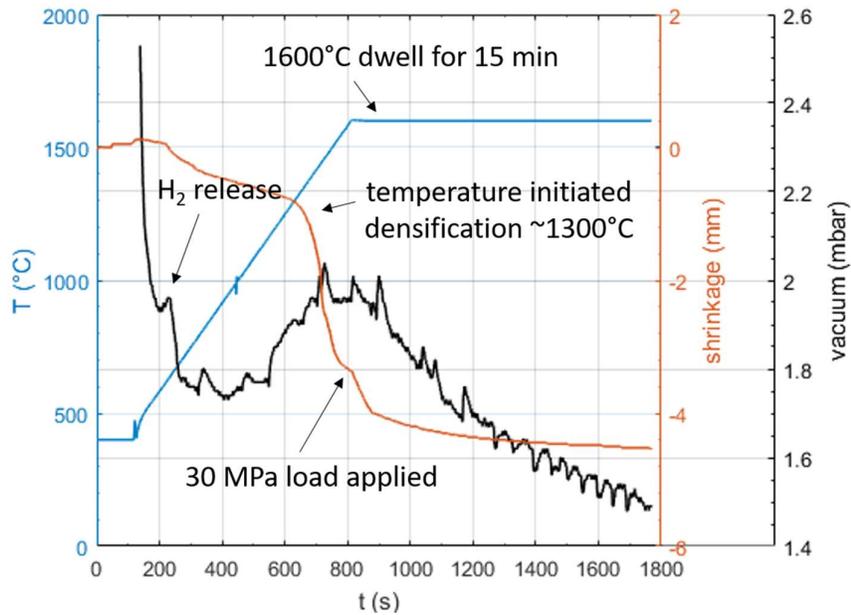

**Fig. 3.** Evolution of temperature, powder compact shrinkage, and vacuum levels during a typical SPS cycle to higher temperatures.

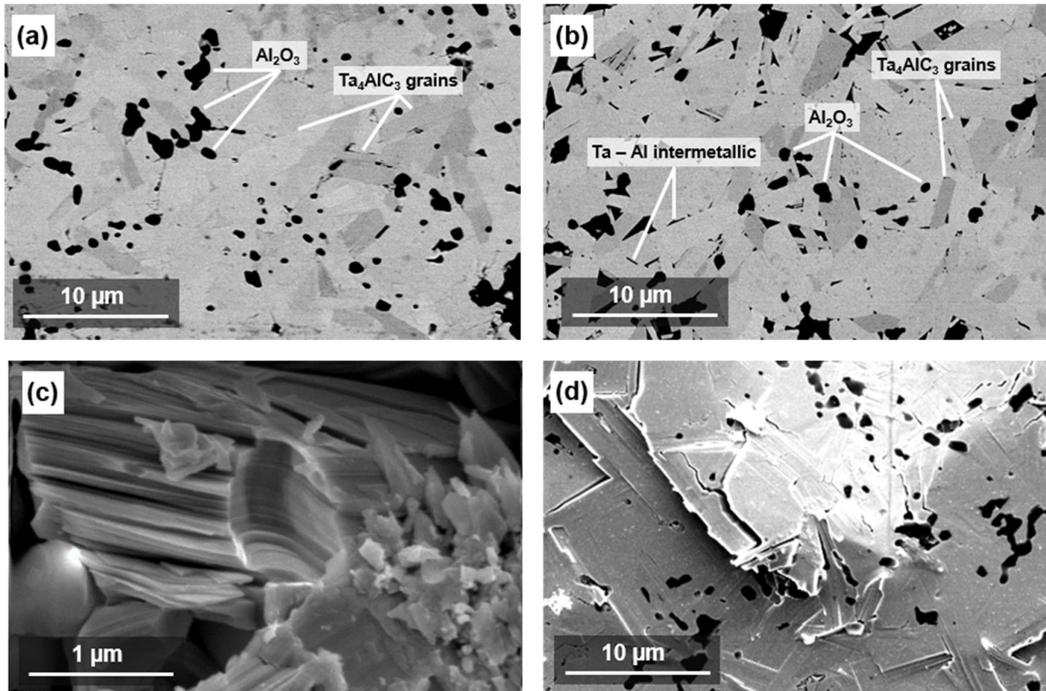

**Fig. 4.** Backscattered electron (BSE) images of $Ta_4AlC_3$ produced at 1500°C by (a) HP and (b) SPS. (c) Fracture surface showing the laminated structure of the $Ta_4AlC_3$ grains. (d) Typical damage near the corner of an indentation produced by a Vickers indenter, showing the dislodging of $Ta_4AlC_3$ MAX phase grains.

The SEM micrographs in Fig. 4 confirm the typical layered structure of the MAX phase grains (Fig. 4c), and show the damage at the corner of an indentation produced by a Vickers indenter (Fig. 4d). The MAX phase grains appear randomly oriented in samples synthesized by both HP and SPS, and the latter showed slightly smaller grains and higher amounts of residual Ta-Al intermetallics. Even though this claim is not supported by a detailed grain size distribution and texturing study, it is a reasonable consequence of the shorter sintering times during SPS.

Dark areas in Figs. 4a and 4b correspond to Ta-Al intermetallics, located between the MAX phase lamellae, and small equiaxed grains of $Al_2O_3$, formed by the aluminothermal reduction of oxide impurities in the starting Ta powder. As mentioned earlier, the fraction of $Al_2O_3$ was estimated around 1 ± 2 wt% by Rietveld refinement, whereas the amount of Ta-Al intermetallic was too small to be detected/quantified by XRD.

### 3.4. Mechanical properties

Table 3 summarizes the measured mechanical properties of the $Ta_4AlC_3$-based samples produced by HP at 1500°C under 30 MPa. The Vickers hardness ($HV_{30}$) was measured to be 6.1 ± 0.4 GPa. This value is higher than the earlier reported 5.1 ± 0.1 GPa for $Ta_4AlC_3$ produced by HP at 1500°C under 25 MPa, starting from metallic Ta powders [6]. The higher hardness of the samples produced in this work could be associated with the dispersion of fine $Al_2O_3$ grains in the material bulk.

**Table 3**
Measured mechanical properties of a sample with 99 wt% α-$Ta_4AlC_3$, 93.6% TD, and an average (grain length = 6 ± 1 μm, grain width = 1.7 ± 1 μm).

| Young's modulus (GPa) | Flexural strength (MPa) | Vickers Hardness (GPa) | SEVNB $K_{Ic}$ (MPa·m$^{1/2}$) |
|---|---|---|---|
| 248 ± 27 | 407 ± 50 | 6.1 ± 0.4 | 5.0 ± 0.6 |

The RT Young's modulus was 248 ± 27 GPa, which is value is significantly lower than the earlier reported value of 324 GPa [6]. This observed difference in elastic stiffness is not perfectly understood, but could be partly attributed to the higher porosity in the samples produced in this work. The flexural strength of $Ta_4AlC_3$ was measured to be 407 ± 50 MPa, which is comparable to the reported 372 ± 20 MPa, for a material with an average grain length of 10 μm and width of 3 μm [6]. The slight strengthening effect can be explained by the finer average grain size of the samples discussed in this work. The SEVNB fracture toughness was found to be 5.0 ± 0.6 MPa·m$^{1/2}$, which is lower than the previously reported 7.7 ± 0.5 MPa·m$^{1/2}$ [6]. Reduction of the material's fracture toughness with reduced grain size is commonly observed in the MAX phases [1]. It should, however, be noted that the herein reported $K_{Ic}$ values correspond to non-fully-dense $Ta_4AlC_3$ samples, where the residual porosity is expected to have decreased the overall fracture toughness.

### 3.5. Thermal decomposition in vacuum

Heat treatments were performed at different time-temperature combinations under a $10^{-5}$ mbar vacuum on polished α-$Ta_4AlC_3$ discs with closed porosity (density: 85% of the TD of α-$Ta_4AlC_3$); these discs were produced by HP at 1500°C under 7 MPa. A SEM image of the rather porous, untreated material is shown in Fig. 5a. A material with higher starting porosity was used on purpose in order to facilitate the evaporation of Al from the bulk of the degrading MAX phase. Fig. 5b shows the morphology of a $Ta_4AlC_3$ sample heat-treated at 1250°C for 33 h. Large coalesced pores formed, a fact that is clearly visible by comparing the microstructures before (Fig. 5a) and after (Fig. 5b) annealing, respectively.

After thermal decomposition, all discs exhibited a rough and compact superficial scale, which was identified by XRD as cubic $TaC_x$. A similar scale was found on $Ta_2AlC$ after annealing in Ar above 1650°C [17]. The stoichiometry was determined to be $TaC_{0.89}$, using the correlation $a(x) = 4.3007 + 0.1563x$, where $a = 4.43(9)$ Å is the lattice parameter of cubic $TaC_x$ and x is the carbon content [30].

An overview of a wider section near the surface of a sample annealed at 1250°C for 33 h is shown in Fig. 6a: one may discern the $TaC_x$ surface layer (Fig. 6b), a transition zone containing a biphasic material (Fig. 6c), and the pristine MAX phase material furthest away from the surface. The large difference in material removal rates during polishing of the metallographic cross-section suggests a substantial hardness difference between the affected and pristine regions. The decomposed layer is separated from the pristine MAX phase ceramic through a transition zone that can be identified by a gradual change in compositional contrast (Fig. 6a). The thickness of the decomposed layer was determined in all heat treatments by measuring the position of the decomposition front (i.e., the end of the decomposed layer) by a sudden drop in Al content, as shown in the EDS line scan of Fig. 6a.

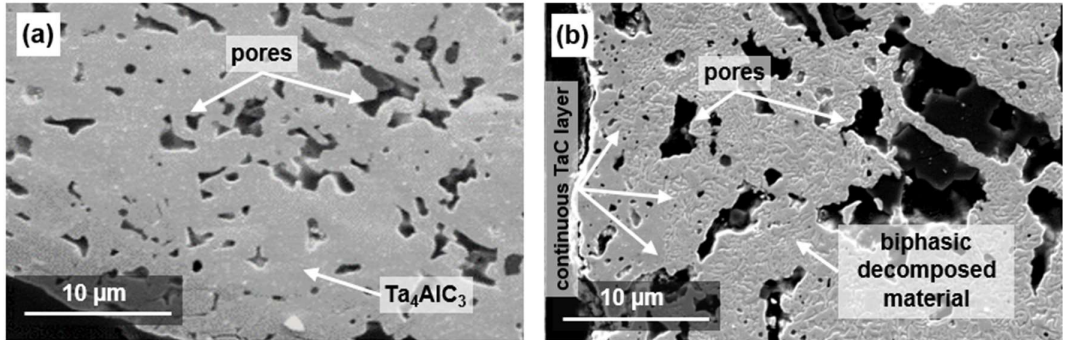

**Fig. 5.** Secondary electron (SE) images of 85% dense $Ta_4AlC_3$ (a) before and (b) after thermal decomposition at 1250°C for 33 h.

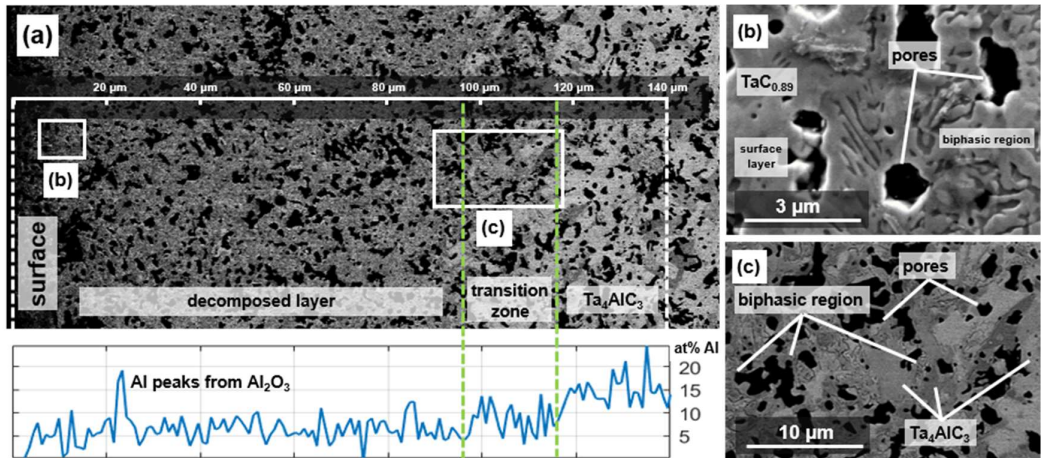

**Fig. 6.** (a) BSE micrograph of a near-surface section of $Ta_4AlC_3$ treated at 1250°C for 33 h and corresponding EDS line scan showing Al concentration. (b) SE image of the decomposed layer near the specimen surface. (c) BSE image of the MAX phase decomposition front, showing the biphasic region and the original, elongated $Ta_4AlC_3$ grains.

When the thickness of the decomposed layer is plotted vs. $t^{1/2}$, straight lines passing through the origin are obtained; the slope of these lines is temperature-dependent (Fig. 7a). These results suggest a parabolic growth rate of the decomposed layer at 1200°C and 1250°C, implying diffusion-controlled kinetics. The parabolic decomposition rate constants were estimated to be 0.023 µm$^2$/s at 1200°C and 0.096 µm$^2$/s at 1250°C. Fig. 7b shows the thin layer obtained after annealing at 1250°C for only 30 min. The continuous TaC$_x$ surface layer (shown in Figs. 5b and 6a) is not present after 30 min at 1250°C.

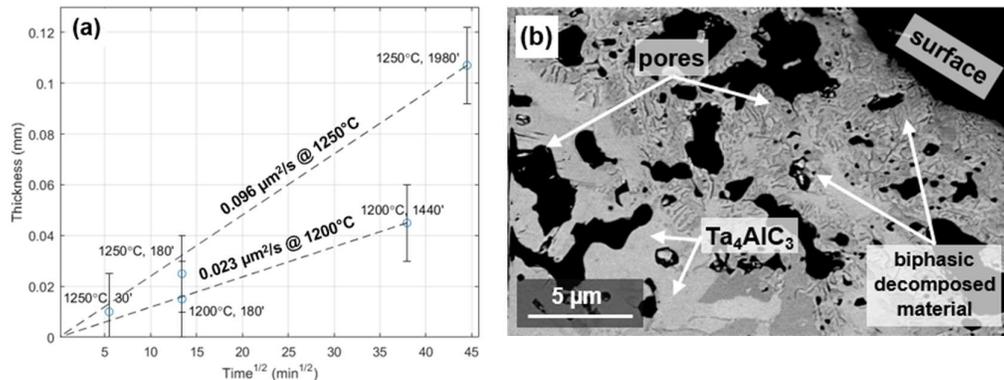

**Fig. 7.** (a) Parabolic plot of the thickness of the decomposed layer as function of the annealing time and temperature. (b) BSE image of Ta$_4$AlC$_3$ annealed at 1250°C for 30 min, showing a considerably thinner decomposed layer without a TaC$_x$ surface layer.

The observed parabolic decomposition rate agrees with similar findings on a similar experiment, where Ti$_3$SiC$_2$ was heat treated in contact with graphite, allowing C to diffuse inward and carburize the material, producing TiC$_x$ [14].

Detailed investigations of the biphasic region were conducted by means of TEM bright-field imaging and SAED on FIB foils extracted from a specimen annealed at 1250°C for 33 h. EDS and WDS elemental point analysis could not quantify the C content of the two distinct phases precisely; however, the overall information acquired by TEM suggested that the two phases were cubic TaC (C/Ta = 0.975, elemental contents in at%; darker Z-contrast phase) and hexagonal α-TaC$_{0.5}$ (C/Ta = 0.44, elemental contents in at%; brighter Z-contrast phase). SAED patterns confirming the formation of cubic TaC and α-TaC$_{0.5}$ are included in Fig. 8: Fig. 8b is a SAED pattern of α-TaC$_{0.5}$ and Figs. 8c-8d are SAED patterns of cubic TaC. Fig. 8a shows a dark-field STEM image of the area in the biphasic region that provided the SAED patterns.

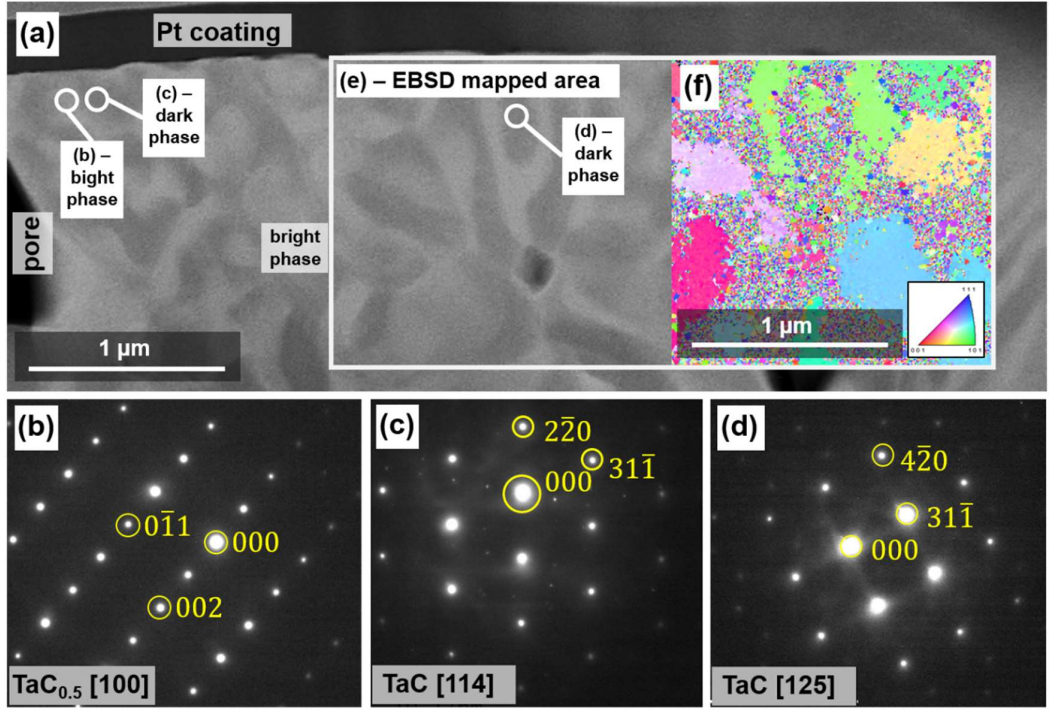

**Fig. 8.** (a) Dark-field STEM image of the biphasic region. (b) SAED pattern of α-TaC$_{0.5}$ along the [100] zone axis. (c) SAED pattern of TaC along the [114] zone axis. (d) SAED pattern of TaC along the [125] zone axis. (e,f) Dark-field STEM image and EBSD orientation map of an ares in the biphasic region.

Cubic TaC was positively identified by EBSD, as shown in the grain orientation map of Fig. 8f over the area shown in Fig. 8e, while EBSD was unable to detect the α-TaC$_{0.5}$ phase. The large difference in the grain size of these two phases could be attributed to large differences in the undercoolings (i.e., ΔT below the melting point) driving their formation in the 1200-1250°C range. According to the Ta-C equilibrium phase diagram (Fig. 9a [30]), the melting point of TaC$_y$ is appreciably higher than the melting point of α-TaC$_{0.5}$; hence, the undercooling driving the crystallisation of TaC$_y$ is higher than that driving the crystallisation of α-TaC$_{0.5}$. Based on these considerations, it is reasonable to expect that the size of cubic TaC grains is larger than that of α-TaC$_{0.5}$ grains, in agreement with the findings of this work. The crystallisation of both phases is presumably assisted by elemental diffusion processes facilitated by the presence of an Al$_3$Ta$_{1-x}$ liquid phase, which forms in the 1200-1250°C range of the annealing treatments performed in this work and in agreement with the Al-Ta phase diagram (Fig. 9c [31]). A bimodal distribution of coarser TaC grains and much finer α-TaC$_{0.5}$ grains has been previously reported [31], as the result of solidification from a Ta$_x$C$_{1-x}$ melt; in that study, the difference in the grain size of the two phases was also attributed to the large difference in the undercoolings of the higher melting point TaC phase and the lower melting point α-TaC$_{0.5}$ phase.

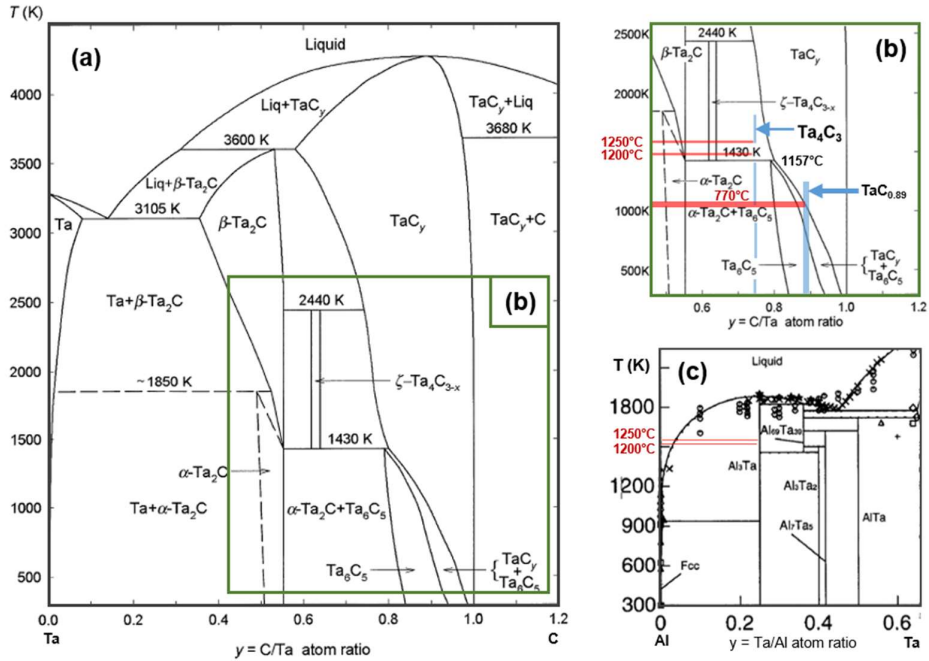

**Fig. 9.** (a) Equilibrium phase diagram of the Ta-C system (adapted from [30]) and (b) detail in the area of interest for this work. (c) Equilibrium phase diagram of the Al-Ta system (adapted from [32]).

The exact decomposition mechanism can only be elucidated by detailed in-situ studies. Since there is no proof that Al is expelled directly in the gaseous form, $Ta_4AlC_3$ might as well decompose initially into a binary carbide and a Ta-Al intermetallic according to reaction (8). The appearance of the transition zone in Figs. 6c and 7b suggests that this decomposition consumes gradually the $Ta_4AlC_3$ grains. When the decomposition occurs at a temperature in the 1200-1250°C range, the $Al_3Ta$ intermetallic is in equilibrium with a liquid Al-Ta phase, as shown in the Al-Ta phase diagram of Fig. 9c [32]. The Al readily evaporates from this liquid phase, either at the surface or through the pore network, according to reaction (9). The produced Ta and $TaC_x$ react with each other and the expected overall composition is $Ta_4C_3$. According to the Ta-C equilibrium phase diagram (Figs. 9a and 9b [33]), $Ta_4C_3$ falls within the two-phase field of cubic $TaC_{1-x}$ and trigonal $\zeta$-$Ta_4C_{3-x}$ above 1157°C. Below 1157°C, the Ta-C system is characterised by a two-phase field at the exact $Ta_4C_3$ stoichiometry, which comprises the hexagonal $\alpha$-$TaC_{0.5}$ and cubic $Ta_6C_5$ phases.

Although the $Ta_6C_5$ phase was predicted by ab initio calculations [34], it was not directly observed in experimental studies on Ta-C ceramics produced by hot isostatic pressing [31]. This phase was also not observed in the present work, as all SAED patterns matched the cubic TaC.

The absence of the $\zeta$-$Ta_4C_{3-x}$ phase led to the conclusion that the system reached equilibrium in the $\alpha$-$TaC_{0.5}$/TaC region during cooling, as suggested by reaction (10). Specifically, sub-stoichiometric $TaC_{0.89}$ corresponds to an equilibrium temperature of roughly 770°C. At this temperature, the expected molar ratio calculated by the lever rule is 64.1 at% for $TaC_{0.89}$ and 35.9 at% for $\alpha$-$TaC_{0.5}$. Given the similar density of the two phases, these ratios agree with the areal distribution of the two phases in micrographs of the biphasic region. Should the decomposition occur by simple basal plane diffusion of Al and vapour pressure-driven Al evaporation from the pristine $Ta_4AlC_3$, the overall reaction would be (11) and the leftover carbide would possibly stabilize in its biphasic form by spinodal decomposition, as suggested by the microstructure presented in Fig. 8a. The possible reaction scenarios proposed above are summarized by the following reactions:

$$Ta_4AlC_3 \rightarrow (3/x)TaC_{0.75 \leq x \leq 1} + AlTa_{4-3/x} \qquad (8)$$
$$AlTa_{4-3/x} \rightarrow (4-3/x)Ta + Al_{(g)} \qquad (9)$$
$$(1/4)(4-3/x)Ta + (1/4)(3/x)TaC_{0.75 \leq x \leq 1} \rightarrow (0.359)\alpha\text{-}TaC_{0.5} + (0.641)TaC_{0.89} \qquad (10)$$
$$(1/4)Ta_4AlC_3 \rightarrow (0.359)\alpha\text{-}TaC_{0.5} + (0.641)TaC_{0.89} + (1/4)Al_{(g)} \qquad (11)$$

The formation of porosity is a direct consequence of the fact that both produced phases have higher densities than the starting MAX phase. A similar shrinkage and pore formation mechanism was described for the decomposition of $Ti_3SiC_2$ thin films in vacuum at 1100–1200°C [16].

Pore coalescence occurring during the decomposition process showed the physical limits of producing $TaC_x$ ceramics with controlled-porosity by annealing bulk $Ta_4AlC_3$ samples. In order to achieve homogeneous and microporous $TaC_x$ samples for ISOL targets, future studies might focus on enhancing Al evaporation by annealing $Ta_4AlC_3$ powder compacts or porous preforms, as opposed to the bulk samples used in this study. Higher temperature regimes might also be explored.

## 4. Conclusions

A powder metallurgical route was adopted to produce bulk, predominantly single-phase $Ta_4AlC_3$ MAX phase ceramics by HP or SPS starting from $Ta_2H/Al/C$ powder mixtures. Achieving a high $\alpha\text{-}Ta_4AlC_3$ phase purity required the refinement of the Ta starting powder below 10 μm by hydrogenation and milling.

Regarding the reactive sintering technique, HP at 1500°C or 1600°C at 30 MPa produced phase-pure $\alpha\text{-}Ta_4AlC_3$. SPS also proved to be a valid technique for the production of this MAX phase, although the possibility to operate with fast heating/cooling cycles and short dwell times proved counterproductive in terms of producing phase-pure $\alpha\text{-}Ta_4AlC_3$. By using SPS at 1600°C, a fraction of the $\beta\text{-}Ta_4AlC_3$ allotrope formed, whereas using SPS at 1650°C resulted in the formation of binary $TaC_x$ as a result of the dissociation of the $Ta_4AlC_3$ phase.

The $Ta_4AlC_3$ sample with the highest phase purity was produced by HP at 1500°C and had the following properties: flexural strength: 407 ± 50 MPa, SEVNB toughness: 5 ± 0.6 MPa·m$^{1/2}$, Young's modulus: 248 ± 27 GPa, and Vickers hardness, $VH_{30}$: 6.1 ± 0.4 GPa. The thermal decomposition of $Ta_4AlC_3$ at 1200°C and 1250°C in high vacuum ($10^{-5}$ mbar) was studied with the aim of characterizing the MAX phase decomposition products and associated kinetics. A biphasic carbide region of enhanced porosity resulted from the decomposition of $\alpha\text{-}Ta_4AlC_3$, with a parabolic decomposition rate constant of 0.023 μm$^2$/s and 0.096 μm$^2$/s at 1200°C at 1250°C, respectively. Since such thermal decomposition could be initiated several hundred degrees below the formation temperature of $Ta_4AlC_3$, it is concluded that this material is not stable at high temperatures and in high vacuum.

**Acknowledgements**

M. Griseri thanks SCK•CEN for his PhD fellowship through the ISOL@MYRRHA project. Additional acknowledgments are due to the technical and academic staff of the Materials Engineering Department (MTM) of the KU Leuven for practical help and support. B. Tunca acknowledges the financial support of the SCK•CEN Academy for Nuclear Science and Technology. The authors acknowledge the Hercules Foundation under Project AKUL/1319 (CombiS(T)EM). M. W. Barsoum acknowledges funding from the Euratom research and training programme 2014–2018 under grant agreement No. 740415 (H2020 IL TROVATORE).